\documentclass[11pt, oneside]{article}   	
\usepackage{geometry}                		
\usepackage{graphicx}				
\usepackage{amssymb}
\usepackage{amsmath}

\title{Simplified Hartree-Fock Computations on Second-Row Atoms}
\author{S. M. Blinder\\Wolfram Research, Inc.\\Champaign, IL 61820, USA}

\begin{document}
\maketitle

Modern computational quantum chemistry has developed to a large extent beginning with applications of the Hartree-Fock method to atoms and molecules[1].
Density-functional methods have now largely supplanted conventional Hartree-Fock computations in current applications of computational chemistry. Still, from a pedagogical point of view, an understanding of the original methods remains a necessary preliminary. However, since more elaborate Hartree-Fock computations are now no longer needed, it will suffice to consider a computationally simplified application of the method[2]. Accordingly, we will represent the many-electron wavefunction by a single closed-shell Slater determinant and use basis functions which are simple Slater-type orbitals. This is in contrast to the more elaborate Hartree-Fock computations, which might use multi-determinant wavefunctions and double-zeta basis functions. All of the symbolic and numerical operations were carried out using Mathematica.

We consider Hartree-Fock computations on the ground states of the second-row atoms He through Ne, $Z=2$
to 10, using the simplest set of orthonormalized $1s$, $2s$ and $2p$  orbital functions:
\begin{eqnarray}
\psi_{1s}=\frac{\alpha^{3/2}}{\sqrt\pi} e^{-\alpha r},\qquad
 \psi_{2s}=\sqrt{\frac{3\beta^5}{\pi(\alpha^2-\alpha\beta+\beta^2)}} \left(1-\frac{\alpha+\beta}{3}r\right) e^{-\beta r},
 \nonumber  \\   \psi_{2p\{x,y,z\}}=\frac{\gamma^{5/2}}{\sqrt\pi} r e^{-\gamma r} \{\sin\theta \cos\phi, \sin\theta \sin\phi,
 \cos\theta\}. \qquad
\end{eqnarray}
An orbital function multiplied by a spin function $\alpha$ or $\beta$ is known as a spinorbital, 
a product of the form
\begin{equation}
\phi(x)=\psi({\bf r})\begin{cases}
      \alpha &  \\
      \beta  &
    \end{cases}.
\end{equation}

A simple representation of a many-electron atom is given by a Slater determinant constructed from $N$ occupied spin-orbitals:\begin{equation}
\Psi(1, \dots, N)=\frac{1}{\sqrt{N!}} \begin{vmatrix} \phi_a(1) & \phi_b(1) & \dots & \phi_n(1) \\ 
\phi_a(2) & \phi_b(2) & \dots & \phi_n(2) \\  \vdots & \vdots & \ddots & \vdots \\ \phi_a(N) & \phi_b(N) & \dots & \phi_n(N)
\end{vmatrix},
\end{equation}
where the subscripts $a, b, \dots$ stand for spinorbitals, for example, $1s\alpha, 2p_x\beta$, etc. and the arguments 1, 2, $\dots$ are abbreviations for $x_1, x_2, \dots$. 

The Hamiltonian for an $N$-electron atom, in hartree atomic units, is given by:
\begin{equation}
H=\sum_{i=1}^N \left\{-\frac{1}{2} \nabla_i^2-\frac{Z}{r_i}\right\} +\sum_{i> j}^N \frac{1}{r_{ij}}.
\end{equation}
The corresponding approximation to the total energy is then given by
\begin{equation}
E=\langle\Psi|H|\Psi\rangle =        \sum_{i=1}^N H_i + \sum_{i> j}^N (J_{ij}-K_{ij}),
\end{equation}
where $H$, $J$ and $K$ are, respectively, the core, Coulomb and exchange integrals:
\begin{equation}
H_i = \int d^3{\bf r} \  \psi_i^*({\bf r}) \left\{-\frac{1}{2} \nabla^2  -\frac{Z}{r}\right\} \psi_i({\bf r}),
\end{equation}
\begin{equation}
J_{ij} = \int\int  d^3{\bf r} d^3{\bf r'} \ \frac{|\psi_i({\bf r})|^2 |\psi_j({\bf r'})|^2}{|{\bf r}-{\bf r'}|},
\end{equation}
\begin{equation}
K_{ij}=
 \int\int d^3{\bf r} \ d^3{\bf r'} \  \psi_i^*({\bf r}) \psi_j^*({\bf r'}) \frac{1}{|{\bf r}-{\bf r'}|} 
 \psi_i({\bf r'}) \psi_j({\bf r}) \langle \sigma_i | \sigma_j \rangle.
\end{equation}
The exchange integral vanishes if 
$\sigma_i \neq \sigma_j$, in other words, if spinorbitals $i$ and $j$ have opposite spins, $\alpha, \beta$ or $\beta, \alpha$.  

Explicit formulas for these integrals are given in the Appendix. Let us illustrate with a couple of examples. The core integral $H_{2p}$ is found from
\begin{equation}
\int_0^\infty \int_0^\pi \int_0^{2\pi}   \frac{\gamma^{5/2}}{\sqrt\pi} r e^{-\gamma r} \cos\theta
\Bigg[ \left\{-\frac{1}{2} \nabla^2  -\frac{Z}{r}\right\} \frac{\gamma^{5/2}}{\sqrt\pi} r e^{-\gamma r} \cos\theta \Bigg] \,
 r^2  \sin\theta \, dr\,d\theta\,d\phi,
\end{equation}
using the Laplacian in spherical polar coordinates
\begin{equation}
\nabla^2=\frac{1}{r^2} \frac{\partial}{\partial r} r^2  \frac{\partial}{\partial r}
+\frac{1}{r^2\sin\theta}  \frac{\partial}{\partial \theta} \sin\theta  \frac{\partial}{\partial \theta}
+\frac{1}{r^2\sin^2\theta}  \frac{\partial^2}{\partial \phi^2}.
\end{equation}
The symbolic computation can be carried out using Mathematica, giving the result
\begin{equation}
H_{2p}=\frac{1}{2}(\gamma^2-Z \gamma).
\end{equation}

The Coulomb integral $J_{1s,2p}$ is found from
\begin{equation}
(2\pi)^2\int_0^\infty \int_0^\pi\int_0^\infty \int_0^\pi \, \psi_{1s}(r_1)^2 \, \frac{1}{r_{12}}
\psi_{2pz}(r_2,\theta_2)^2\,
 r_1^2  \sin\theta_1 \, dr_1\,d\theta_1 \, r_2^2  \sin\theta_2 \, dr_2\,d\theta_2
\end{equation}
The interelectronic potential can be expanded in a series of Legendre polynomials:
\begin{equation}
\frac{1}{r_{12}}=\sum_{l=0}^\infty \frac{r_<^l}{r_>^{l+1}} P_l(\cos\Theta_{12}),
\end{equation}
where $r_>$ and $r_<$ are the greater and lesser of $r_1$ and $r_2$. In terms of individual particle coordinates
\begin{equation}
 P_l(\cos\Theta_{12})=\frac{4\pi}{2l+1} \sum_{m=-l}^l Y_l^{m}(\theta_1,\phi_1)^* \, Y_l^{m}(\theta_2,\phi_2).
\end{equation}
In the integral (12), only the $l=0$ term contributes.
We must consider  the contributions both with $r_2 > r_1$ and  $r_1 > r_2$. Using Mathematica, the integral evaluates to
\begin{equation}
J_{1s,2p}=
\frac{\alpha \gamma(\alpha^4 + 5 \alpha^3 \gamma + 
   10 \alpha^2 \gamma^2 + 10 \alpha \gamma^3 + 
   2 \gamma^4)}{2 (\alpha+ \gamma)^5}.
\end{equation}

The energy $E$ is  optimized by seeking a minimum as a function of the orbital parameters $\alpha$, $\beta$ and $\gamma$. For example, for $Z=6$, the carbon atom with ground state configuration $1s^2 2s^2 2p \, 2p' \ {}^3 {\rm P}_0$, we have
\begin{eqnarray}
E(\alpha,\beta,\gamma)=2H_{1s}+2H_{2s}+2H_{2p}+J_{1s,1s}+J_{2s,2s}+4J_{1s,2s}-2K_{1s,2s}+ \nonumber \\
4J_{1s,2p}-2K_{1s,2p}+4J_{2s,2p}-2K_{2s,2p}+J_{2p,2p'}-K_{2p,2p'}
\end{eqnarray}

Following is a tabulation of the results, showing the optimal parameters and the calculated energies. Energies are expressed in atomic units. The computations made use of the {\tt FindMinimum} function in Mathematica. Each computation took  approximately 0.005 seconds of CPU time (using a MacBook Pro with the Apple M1 chip).
For comparison, we also include the results of the best Hartree-Fock computations[3] and the exact nonrelativistic energies of the atomic ground states. Our approximate energies are within 1\% of the accurate H-F values, obtained with an order of magnitude less computational effort. \newpage

\section*{\centerline {Hartree-Fock Results}}
\begin{center}
\begin{tabular}{|c|c|c|c|c|c|c|c|c|c|}
 \hline
  $Z$ & atom & configuration & $\alpha$ & $\beta$ & $\gamma$ & calculated $E$ & best H-F & exact  \\ \hline 
   2 & He & $1s^2 \ {}^1$S & 1.6875 & &  &  -2.84766 & -2.86168 & -2.903385  \\ \hline
    3 & Li & $1s^22s \ {}^2$S & 2.69372 & 0.766676 &  & - 7.41385 & -7.43273 & -7.477976  \\ \hline 
    4 & Be & $1s^22s^2 \ {}^1$S  & 3.70767 & 1.15954 &  & - 14.5300 & -14.5730 & -14.668449  \\ \hline 
    5 & B & $1s^22s^22p \ {}^2$P  & 4.71099 &  1.57921 &  1.18716 & -24.4506 & -24.5291 & -24.658211  \\ \hline 
    6 & C & $1s^22s^22p^2 \ {}^3$P  & 5.71244 &  1.98775 &  1.51874 & -37.5471 &  -37.5471 & -37.855668  \\ \hline 
    7 & N & $1s^22s^22p^3 \ {}^4$S  & 6.71293 &  2.39148 & 1.84551 & -53.9624 &   -54.4009 & -54.611893  \\ \hline
     8 & O & $1s^22s^22p^4 \ {}^3$P  & 7.71286 & 2.79267 &  2.16972 & -74.1624 &   -74.8094 & -75.109991  \\ \hline 
    9 & F & $1s^22s^22p^5 \ {}^2$P  & 8.71243 & 3.19236 &   2.49238 & -98.3972 &   -99.4093 & -99.803888  \\ \hline 
    10 & Ne & $1s^22s^22p^6  \ {}^1$S  & 9.71176 & 3.59108 &  2.81404 &-126.971 &   -128.547 & -128.830462  \\ \hline  
\end{tabular}
\end{center}

\includegraphics[height=8cm]{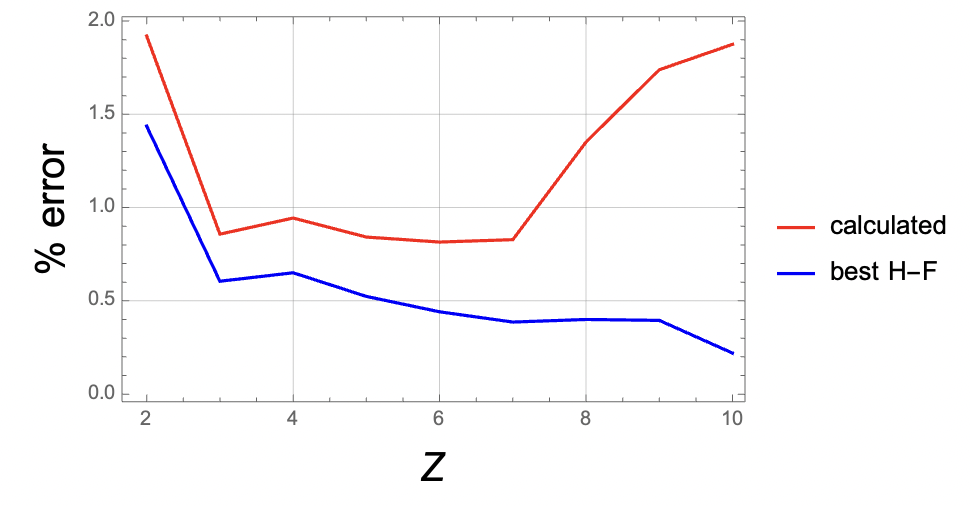}

 \newpage

\section*{\centerline {Appendix: Core, Coulomb and Exchange Integrals}}
\includegraphics[height=12cm]{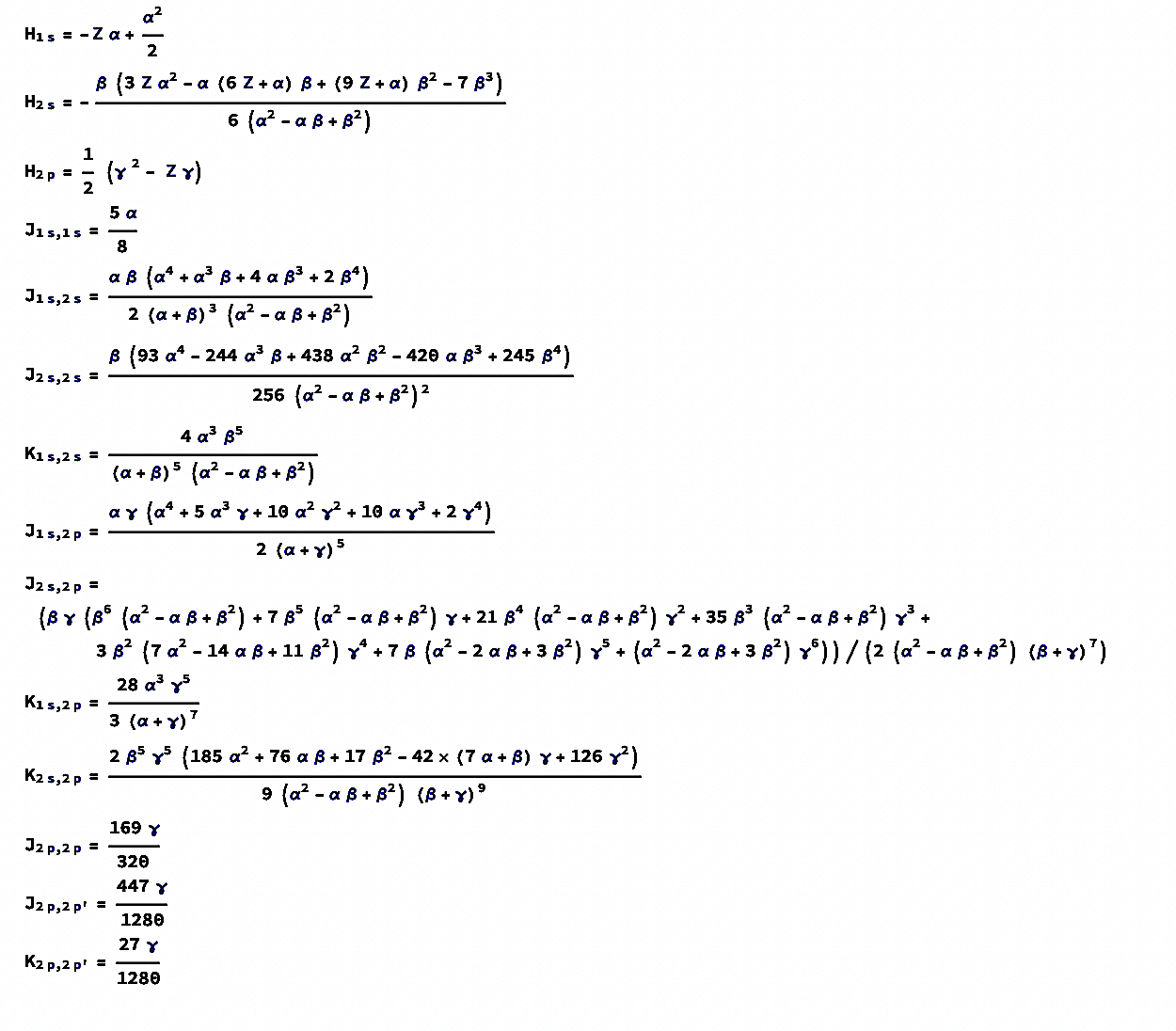}

\section*{References}
[1] S. M. Blinder, ``Introduction to the Hartree-Fock Method,'' in {\it Mathematical Physics in Theoretical Chemistry} (S. M. Blinder and J. E. House, eds.), Elsevier, 2018.

[2] This paper is based in the Wolfram Demonstration: \\
https://demonstrations.wolfram.com/SimplifiedHartreeFockComputationsOnSecondRowAtoms/

[3] Atomic Data and Nuclear Data Tables,  {\bf 95}(6), 2009 pp. 836-870.

\end{document}